\documentclass{pasj00}

\begin{document}
\SetRunningHead{K. Murata et al.}{PAH, Br$\alpha$, and LIR of local galaxies observed with {\it AKARI}}
\Received{2017/04/03}
\Accepted{2017/06/27}

\title{Relation between polycyclic aromatic hydrocarbon, Br$\alpha$ and infrared luminosity of local galaxies observed with AKARI}

\author{Kazumi \textsc{Murata},
  Takao \textsc{Nakagawa},
  Hideo \textsc{Matsuhara}, and
  Kenichi {\sc Yano}
  }
  \affil{Institute of Space and Astronautical Science, Japan Aerospace Exploration Agency, Sagamihara, Kanagawa 252-5210, Japan}
  \email{murata@ir.isas.jaxa.jp}


\maketitle

\begin{abstract}
  We produce a catalogue of polycyclic aromatic hydrocarbon (PAH) 3.3 $\mu$m, Br$\alpha$ and infrared luminosity ($L$(IR)) of 412 local galaxies, and investigate a relation between these physical parameters. 
We measure the PAH 3.3 $\mu$m and Br$\alpha$ flux using {\it AKARI} 2-5 $\mu$m spectra and the $L$(IR) using the {\it AKARI}-all-sky-survey data. 
The $L$(IR) and redshift ranges of our sample are $L$(IR)=$10^{9.7-12.8}$L$_\odot$ and $z_{\rm spec}=0.002-0.3$, respectively.
We found that the ratio of $L$(PAH 3.3 $\mu$m) to $L$(IR) is constant at $L$(IR) $<$ $10^{11} \rm L_\odot$ whereas it decreases with the $L$(IR) at higher $L$(IR). 
Also, the ratio of $L$(Br$\alpha$) to $L$(IR) decreases with the $L$(IR).
The both $L$(PAH)/$L$(IR) and $L$(Br$\alpha$)/$L$(IR) ratios are not strongly dependent on galaxy type and dust temperature. 
The relative weakness of the two ratios could be attributed to destruction of PAH, a lack of UV photons exciting PAH molecules or ionising hydrogen gas, extremely high dust attenuation, or active galactic nucleus contribution to the $L$(IR). 
Although we cannot determine the cause of the decreases of the luminosity ratios, a clear correlation between them implies that they are related with each other.
The catalogue presented in our work will be available at the {\it AKARI} archive web page.

\end{abstract}

\section{Introduction}
For understanding galaxy evolution, star formation rate (SFR) is one of the most basic parameters of a galaxy.
The SFR of galaxies has been derived by many researchers with various methods. 
Among various SFR tracers, rest frame UV and optical emission have been widely used for estimating SFR, even for high-redshift galaxies. 
Nonetheless, these tracers are strongly attenuated by interstellar dust, which is more serious for galaxies with more luminous at infrared \citep{Takeuchi10}.
To correctly estimate SFR of such galaxies, we need SFR tracers that are less affected by dust attenuation, i.e. those at infrared wavelength. 

There are three kinds of SFR tracers at infrared: polycyclic aromatic hydrocarbon (PAH) emission, hydrogen recombination lines such as Br$\alpha$ (4.052 $\mu$m), and infrared luminosity ($L$(IR)). 
PAH is a large molecule, exists mainly at a photo-dissociation region (PDR), absorbs UV photons from young stars and emits the energy at near- to mid-infrared such as 3.29, 6.2, 7.7, and 11.3 $\mu$m \citep{Tielens08}. Because the energy source is UV light from young stars, PAH emission reflects the star-forming activities of galaxies. 
%
Br$\alpha$ line is a hydrogen recombination line emitted from an HII region.
It directly traces the ionising photons produced by young stars.
In addition, assuming the galactic attenuation curve, the dust attenuation is merely A(Br$\alpha$) $\sim$ 0.1A(H$\alpha$) owing to the long wavelength. 
These properties make the Br$\alpha$ line an ideal SFR tracer for galaxies with strong dust extinction like ultra luminous infrared galaxies (ULIRGs). 
$L$(IR) is luminosity integrated over infrared spectrum at 8-1000 $\mu$m. It is emitted by interstellar dust that absorbs UV and optical radiation from young stars.  
Hence for galaxies with extremely strong dust attenuation like ULIRGs, the $L$(IR) approximates the bolometric luminosity and becomes an ideal SFR tracer \citep{Kennicutt1998AR}. 

Combining these SFR tracers provides us various galaxy information because the relation between them is different in different physical condition of galaxies.
For example, if PAH particles are destroyed by strong interstellar radiation, they are not proportional to the SFR of galaxies so that $L\rm (PAH)$/$L$(IR) ratio can be used for inferring physical condition of interstellar medium \citep{Nordon12,Murata14}. 
Hydrogen recombination lines from a dusty HII region could be reduced due to a lack of ionising photons \citep{Valdes05,Yano16}.
These phenomena indicate that combining the three SFR tracers enables us to investigate galaxy properties related to the star-forming activity.

The Japanese infrared astronomical satellite {\it AKARI} \citep{Murakami07} can provide these SFR tracers.
A spectroscopic observation at 2-5 $\mu$m wavelength with the Infrared Camera (IRC; \cite{Onaka07}) is capable of measuring PAH 3.3 $\mu$m and Br$\alpha$ emission.
Furthermore, using the all-sky-survey data with the far-infrared surveyer (FIS; \cite{Kawada07}) we can estimate the $L$(IR) of galaxies. 
Some researchers have provided PAH and Br$\alpha$ luminosity of local galaxies using these unique capabilities \citep{Imanishi08,Imanishi10,Kim12,Lee12,Yamada13,Ichikawa14,Castro14}.
For example Imanishi et al.(2008,2010) provided PAH flux of 154 (U)LIRGs, for 67 of which they also provided Br$\alpha$ flux. 
However, a significant amount of local galaxy data is still unpublished.



In this work, we analyse all {\it AKARI}/IRC spectroscopic data and produce a catalogue of 412 local galaxies with PAH 3.3 $\mu$m and Br$\alpha$ flux.
We also estimate and provide infrared luminosity of these galaxies using {\it AKARI}-all-sky-survey data.
Using these data we investigate the relation between the three SFR tracers. 
This paper is organised as the following.
In section 2 and 3, we describe data used in our study and how we estimated the PAH 3.3 $\mu$m, Br$\alpha$, and infrared luminosity.
In section 4, we show our results and present the catalogue.
In section 5, we discuss the relation between the three SFR tracers.
Finally we summarise our study in section 6.
Throughout our work we adopt a cosmology with $\Omega_M$=0.3, $\Omega_\Lambda$=0.7, and $H_0$=70 km s$\rm ^{-1} Mpc^{-1}$.

\section{Data}
\subsection{AKARI/IRC spectroscopy}
We used spectra observed with the {\it AKARI}/IRC grism spectroscopy. 
The IRC has a 1' $\times$ 1' aperture mask, in which the light from galaxies is incident and dispersed with wavelength resolution of $R\sim$120. 
The {\it AKARI} operation is divided into two phases: Phase 1-2, and phase 3. 
In the phase 1-2, liquid helium is used as a cryogen and observation was conducted at 2006 April to 2007 August.
In the phase 3, however, liquid helium is exhausted and the sensitivity is degraded. The observed period of this phase is from 2008 June to 2010 February.
The spectra were recently reduced by the {\it AKARI} data processing and analysis team in a standard manner with the latest pipeline\footnote{The original version of the pipeline was developed in \citet{Ohyama07}.} (Usui et al. in preparation). 
They extracted the spectra within 10 arcsec width in spatial direction from the reduced images. 

\subsubsection{Sample selection and observation programmes}

\begin{table}[ht]
  \caption{Observation programmes used in our catalogue.
    The second column indicates the number of galaxies in our catalogue.
    We have 10 galaxies taken with multiple programmes, leading the sum of the second column is over the number of the catalogued galaxies.
    \label{tab:prog}}
  \centering
\begin{tabular}{ccc}
  \hline
  Programme& \# of galaxies & targets\\
  \hline
  AGNUL& 180 & LIRG \& ULIRG\\
  AMUSE&4 & $F_{24\mu m} > 5$mJy\\
  BRSFR&17 & $L$(IR) $\rm < 10^{11}L_{\odot}$\\
  CLNSL&6 & composite in BPT\\
  COABS&1 & bright Seyfert-2\\
  DTIRC&11 & data check\\
  EGANS&8 & SDSS at $z$=0.1-0.5\\
  GOALS&48 & LIRG\\
  H2IRC&1 & H$_2$ detected\\
  ISBEG&19 & blue early type\\
  MSAGN&82 & {\it AKARI}/mid-IR AGN\\
  MSFGO&11 & {\it AKARI}/mid-IR\\
  NISIG&10 & star-forming\\
  NULIZ&18 & ULIRG\\
  QSONG&3 & QSO\\
  SYDUS&3 & Seyfert 1\&2\\
  \hline
  \end{tabular}
\end{table}

We used sixteen observation programmes for our sample as summarised in table \ref{tab:prog}.  
Here we briefly describe how targets are selected in each programme.
We note that we included only galaxies with noticeable PAH 3.3 $\mu$m emission in the spectra into our sample, and that some galaxies were duplicately observed in different programmes.

\textbf{AGNUL:}
This programme is planned for observing LIRGs and ULIRGs selected with the Bright Galaxy Sample (BGS; \cite{Soifer87,Sanders95}), the revised BGS \citep{Sanders03}, and the IRAS 1 Jy sample \citep{Kim98}. 
Although the target selection is not complete, the completeness is limited only by the sky position of galaxies.
The programme is conducted both in the phase 1-2 and phase 3.

\textbf{AMUSE:}
This programme is for observing galaxies from the 5 mJy unbiased {\it Spitzer} extra galactic survey (5MUSES; \cite{Wu10}).
The galaxies are brighter than 5 mJy at 24 $\mu$m and located in the {\it Spitzer}-first-look-survey and SWIRE fields. 

\textbf{BRSFR:}
This programme is for observing galaxies whose radial velocity is $\sim$7000 km/s and their infrared luminosity is fainter than $10^{11}L_{\odot}$. 

\textbf{CLNSL:}
This programme is for observing composite galaxies\footnote{ In a BPT diagram \citep{BPT}, they located in an intermediate region between Seyfert and HII galaxy regions.} and low ionisation nuclear emission regions (LINERs) selected from infrared galaxies in \citet{Hwang07}.

\textbf{COABS:}
This programme is for observing nuclei of bright Seyfert-2 galaxies to investigate the physical condition of molecular tori.  

\textbf{DTIRC:}
This is a programme for data checks by the IRC data team.
The targets include standard stars and some galaxies. 

\textbf{EGANS:}
This is a programme for observing SDSS galaxies at $z$=0.1-0.5. 
Our sample includes eight galaxies from this programme.

\textbf{GOALS:}
In this programme the sample is based on the Great Observatory All-Sky LIRG Survey (GOALS; \cite{Armus09}). 
This is a non-bias survey for investigating local LIRGs selected from the revised BGS.

\textbf{H2IRC:}
This programme is conducted for observing galaxies with strong emission of molecular hydrogen.

\textbf{ISBEG:}
This programme is for observing early type galaxies with unusually blue colour selected from the SDSS \citep{Lee10}. 
The redshift range of the targets is $z$=0.02-0.1. 

\textbf{MSAGN:}
It is a follow-up programme of sources detected with the {\it AKARI}-mid-infrared-all-sky survey.
Galaxies with $F(9\mu m)/F(Ks) > 2$ and located at $|b| > 30\,\, \rm deg$ were selected as a target \citep{Oyabu11}.

\textbf{MSFGO:}
It is also a follow-up programme of sources detected with the {\it AKARI}-mid-infrared-all-sky survey.
The sources located at galactic latitude of $|b| < 30 \,\, \rm deg$ were selected as a target.

\textbf{NISIG:}
This is a programme for observing star-forming infrared galaxies selected from the catalogue of \citet{Hwang07}. 

\textbf{NULIZ:}
The targets were selected from 320 ULIRGs in \citet{Hwang07}. 
Their ULIRGs were selected by cross-matching the IRAS faint source catalogue with galaxy redshift surveys: the SDSS DR4 \citep{Adelman-McCarthy06},
2dF Galaxy Redshift survey \citep{Colless01},
and 6dF Galaxy survey \citep{Jones04}. 

\textbf{QSONG:}
This is a programme for observing both low and high redshift quasi stellar objects (QSOs; \cite{Kim15,Jun15}).
We used only low redshift QSOs showing PAH 3.3 $\mu$m emission. 

\textbf{SYDUS:}
This is a programme for observing a 12 $\mu$m sample of nearby Seyfert 1 and 2 galaxies. 


\subsection{AKARI FIS all-sky survey}
We used the data from {\it AKARI}-far-infrared-all-sky survey for estimating the $L$(IR) of our sample. 
We used both the bright source catalogue version 2 (BSC2; Yamamura et al. in prep.) and the intensity map\footnote{Both data products are available in the {\it AKARI} archive page.} \citep{Doi15,Takita15}. 
The detection limit of the BSC2 is 0.44 Jy at the Wide-S band (90 $\mu$m) and 3.4 Jy at the Wide-L band (140 $\mu$m).
The spatial resolution is 78 arcsec at the Wide-S and 88 arcsec at the Wide-L bands, respectively \citep{Takita15}. 


\section{Method}

\subsection{Spectrum extraction}
While we basically used spectra provided by the {\it AKARI} data processing and analysis team, we extracted spectra from the provided images for some galaxies because of the following two reasons.
First, some spectra are failed to be extracted due to a wrong position setting. 
Second, even if multiple galaxies were observed in one image, only one spectrum is extracted. 

While the {\it AKARI} team extracted the spectra with a fixed box size of 10 arcsec, we applied a Gaussian fitting to optimise the aperture size for each galaxy.
First we determined the spatial width and position of the spectra with a Gaussian fitting. 
The fitting was repeated along the wavelength direction.
While this leads different width and position in different wavelength, we took median values and fixed them.

Once width and position were fixed, we extracted the spectra using a Gaussian fitting with only one free parameter of flux. 
We visually confirmed that the fitting was correctly conducted.
We combined the 1D spectra of galaxies that were observed multiple times\footnote{Typically, they were observed three times.}. 

\subsection{Flux calibration}
The spectra were scaled to the 3.4 $\mu$m flux from the Wide field Infrared Survey Explorer ({\it WISE}; \cite{Wright10}) all-sky catalogue owing to the two reasons. 

First, the sensitivity of the IRC is known to be changed with the detector temperature, and the current version of the pipeline does not provide correction for the sensitivity variation while the function is expected to be implemented in the next release. 

Second, in case when the size of a galaxy is larger compared to the aperture used for extracting spectra, the aperture correction is needed. 
Among the {\it WISE} magnitudes, we used the {\it w1mag\_8} for smaller galaxies than the {\it w1rsemi} of 30 arcsec and the {\it w1gmag} for larger galaxies.
The {\it w1mag\_8} is a W1 24.75 arcsec radius aperture magnitude while the {\it w1gmag} is a magnitude in an elliptical aperture whose size is determined by the 2MASS extended source catalogue. To convert a magnitude from Vega to AB, we added 2.699 mag\footnote{http://wise2.ipac.caltech.edu/docs/release/allsky/expsup/sec4$\_$4h.html} to both {\it w1mag\_8} and {\it w1gmag}. 

\begin{figure}
 \begin{center}
   \FigureFile(90,){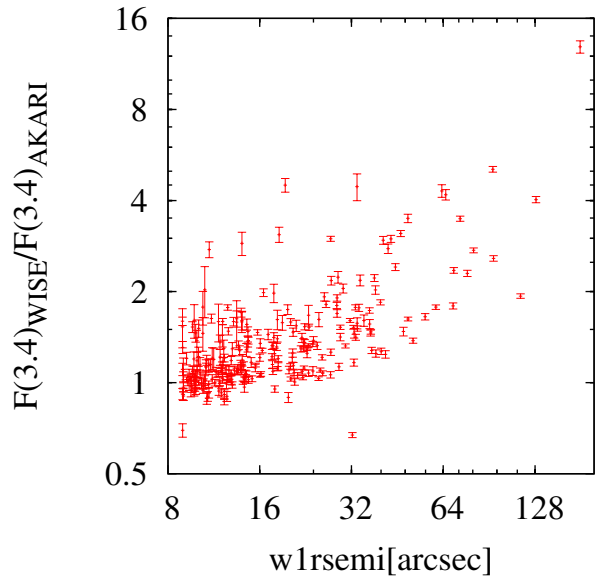} 
 \end{center}
\caption{
  Flux ratio between {\it WISE} and {\it AKARI} against galaxy size, $w1rsemi$.
  The $F(3.4)_{AKARI}$ was estimated from the spectra convolved with the response curve of the {\it WISE} W1 band.
  The sample used in this histogram is from our final catalogue and galaxies rejected from the catalogue are not included (see section \ref{flux}). 
\label{fig:scalehist}}
\end{figure}

Fig.\ref{fig:scalehist} shows the scaling factors applied in our study against galaxy size. 
Despite the scatter, it shows that larger galaxies have a larger scaling factor. 
It also shows that most of our sample have a scale factor of less than 2, and there is only one object with $F_{WISE}/F_{AKARI}$ larger than 10. 

\subsection{Line flux measurements}
\label{flux}
For measuring PAH 3.3 $\mu$m and Br$\alpha$ flux, we performed a Gaussian profile fitting to the spectra.
At first, we determined the baseline with a linear fitting around each line position.
The free parameters are slope and intercept.
We typically used rest frame 3.05-3.2 $\mu$m and 3.4-3.72 $\mu$m for a PAH 3.3 $\mu$m line and 3.9-4.0 $\mu$m and 4.1-4.2 $\mu$m for a Br$\alpha$ line. 
In case S/N ratio at these wavelength range was low, the range was slightly modified, depending on galaxies.  

Second, we fitted each emission line using a Gaussian with the determined baseline. 
The free parameters were flux, central wavelength, and line width. 
The fitted rest frame wavelength range is typically 3.2-3.4 $\mu$m for a PAH 3.3 $\mu$m line and 4.0-4.1 $\mu$m for a Br$\alpha$ line.
Redshifts were estimated from the central wavelength. 
Equivalent widths were measured from the ratio between line and continuum flux.
The properties of these physical values are shown in section \ref{catalogue}.
In case the spectrum show a sub-peak of PAH emission at 3.4 $\mu$m, we avoided it in the fitting processes, following \citet{Imanishi10}.

We included only galaxies whose PAH emission is detected with over 3$\sigma$ into our sample, where we estimated the flux error only from the fitting residual.
As a result, our sample includes 412 galaxies, among which we also detected the Br$\alpha$ emission from 264 galaxies. 

\subsection{Luminosity distance}
The luminosity distances of galaxies were estimated from their redshifts.
The redshifts were obtained from the SIMBAD astronomical database\footnote{http://simbad.u-strasbg.fr/simbad/} or from the NASA/IPAC Extragalactic Database (NED)\footnote{https://ned.ipac.caltech.edu/}.
For galaxies with $z_{\rm spec} < 0.01$ or unknown redshift we used the ``Redshift-Independent Distances'' from the NED instead of distances estimated from the redshift. 
We did not find the redshift-independent distances for nine galaxies at $z_{spec} < 0.01$. 
Among which we estimated the distance of IC0836 ($z_{\rm spec}=0.0092$), NGC5010 ($z_{\rm spec}=0.0099$), and UGC07179 ($z_{\rm spec}=0.0088$) based on their redshift because the individual radial velocity could be neglected.
For the other six galaxies, we cannot estimate their distances. 

\subsection{Infrared luminosity}
We estimated the infrared luminosity of our sample using the bright source catalogue version 2 (BSC2) of the {\it AKARI}-far-infrared-all-sky survey.
We cross-matched the catalogue with our sample using a search radius of 30 arcsec.
Among our sample, 340 galaxies were matched with the BSC2. 
For the remaining galaxies, we performed aperture photometry on the FIS all-sky map \citep{Doi15,Takita15} and 43 galaxies were detected with over 3$\sigma$ at the 90 $\mu$m band.
Most of them are fainter than $F_{\nu}(90) < $ 2 Jy.
We found five bright galaxies that are clearly blended with the next sources.
In total we found 17 galaxies blended in the 90 $\mu$m map, for which we could misidentify or overestimate their flux, and we flagged them in the final catalogue (FISflag in the catalogue; see table \ref{tab:cat}).
We have FIS photometry for 383 galaxies in total.
Among these, the luminosity distance for three galaxies could not be constrained thus we did not derive infrared luminosity for these objects.

To estimate infrared luminosity, we applied an equation suggested in \citet{Solarz16}. 
They derived the equation by comparing the BSC2 with the Infrared Astronomical Satellite (IRAS) all-sky survey. 
The derived equation is as follows.
\begin{eqnarray}
  {\rm log} \, L{\rm(IR)} = 1.016 \, {\rm log} \, L^{\rm{2band}}_{AKARI}/L_\odot + 0.349 \,\,\, (\pm 0.11),
\end{eqnarray}
where,
\begin{eqnarray}
  L^{\rm{2band}}_{AKARI} = \Delta \nu \rm{(Wide-S)} \, L_{\nu}(\rm{90 {\mu}m}) \nonumber \\
  + \Delta \nu \rm{(Wide-L)} \, L_{\nu}(\rm{140 {\mu}m}), 
\end{eqnarray}
\begin{eqnarray}
  \Delta \nu \rm{(Wide-S)}=1.47\times 10^{12}\,\rm{[Hz]},\nonumber\\
  \Delta \nu \rm{(Wide-L)}=0.831\times 10^{12}\,\rm{[Hz]}.\nonumber
\end{eqnarray}
We did not use the 65 $\mu$m band flux because we found that it produces a significant deviation in $L$(IR) for some galaxies. 
For galaxies without 140 $\mu$m band flux, we used only 90 $\mu$m flux to estimate $L$(IR) using the following equation, which was determined to match the $L$(IR) derived above. 
\begin{eqnarray}
  {\rm log} L{\rm(IR)} = 0.976 \,\,  {\rm log} \left[ \, \Delta \nu {\rm (Wide-S)} \, L_{\nu}(\rm{90 {\mu}m})/L_{\odot} \right] \nonumber\\
  + 0.995 \,\,\, (\pm 0.11)
\end{eqnarray}
\par
We added 0.11 dex to the error of the $L$(IR) due to the use of the above equations. 

\section{The PAH 3.3 $\mu$m and Br$\alpha$ flux catalogue}
\label{catalogue}
In this section we provide a catalogue of flux and equivalent width of PAH 3.3 $\mu$m and Br$\alpha$ emission and infrared luminosity. 
We catalogued 412 local galaxies whose PAH 3.3 $\mu$m emission was detected over 3$\sigma$. 
Among our sample, 264 galaxies have Br$\alpha$ flux and equivalent width, and 380 have $L$(IR). 
In our knowledge, it is the largest catalogue of PAH 3.3 $\mu$m and Br$\alpha$ emission. 
The catalogue can be obtained from the {\it AKARI} archive page\footnote{http://www.ir.isas.jaxa.jp/AKARI/Archive/index.html}.
We also present a part of the catalogue in table \ref{tab:cat}. 
We show the properties of our catalogue in the following subsections.

\subsection{Flux and equivalent width}
\begin{figure*}[ht]
   \begin{minipage}{0.49 \hsize}
     \FigureFile(90,){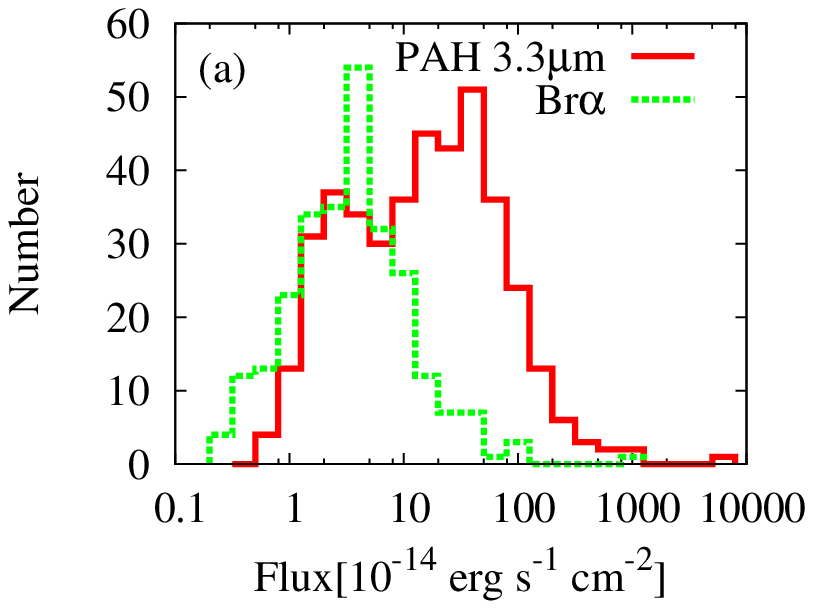}
   \end{minipage}
   \begin{minipage}{0.49 \hsize}
     \FigureFile(90,){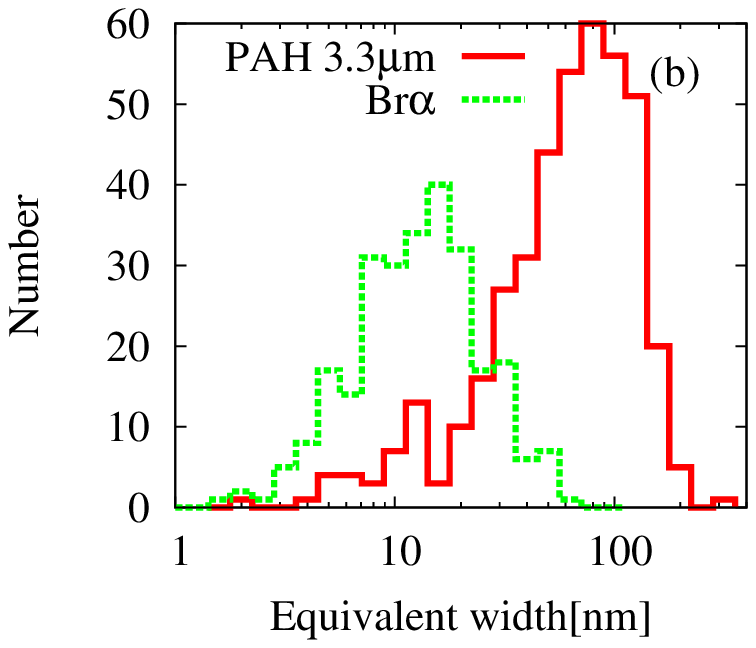}
   \end{minipage}
\caption{Histogram of flux (a) and equivalent width (b) of PAH (solid red line) and Br$\alpha$ (green broken line) emission in our catalogue.
The bin sizes are 0.2 dex for flux histogram and 0.1 dex for those of equivalent width. 
}\label{fig:ewflxhist}
\end{figure*}

Fig.\ref{fig:ewflxhist} shows histograms of flux (a;left) and equivalent width (b;right) of PAH and Br$\alpha$ emission.
From Fig.\ref{fig:ewflxhist}a we can see that most of our sample is distributed at $Flux >$$\rm 1\times 10^{-14} erg s^{-1} cm^{-2}$ for both PAH and Br$\alpha$ emission.

On the other hand, the distribution edge of the equivalent width is different between the two lines; the edge of the EW(Br$\alpha$) is $\rm \sim 10\times$ lower than that of the EW(PAH).
This implies a less bias against low equivalent width, which is clearly different from a narrow-band-imaging-emission-line survey. 


\subsection{Redshift comparison and distribution}
\begin{figure*}[ht]
  \centering
  \begin{minipage}{0.49 \hsize}
    \FigureFile(90,){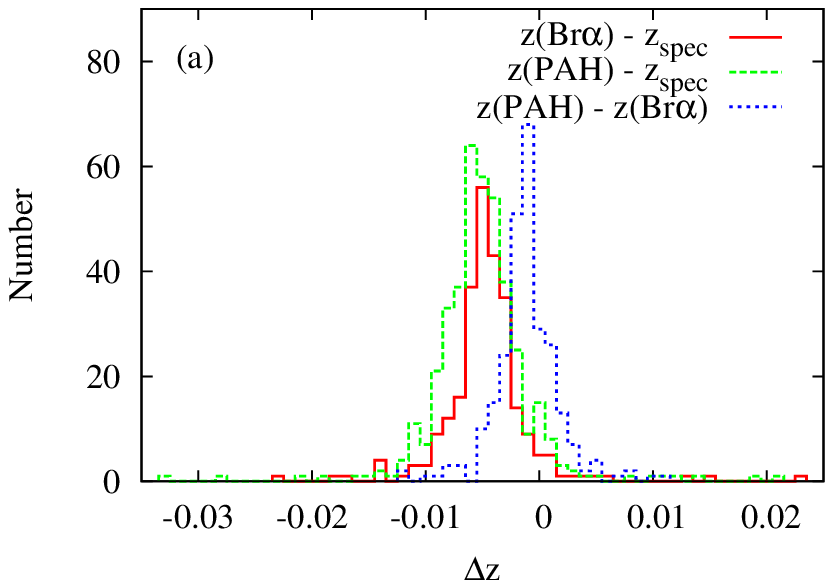}
  \end{minipage}
  \begin{minipage}{0.49 \hsize}
    \FigureFile(90,){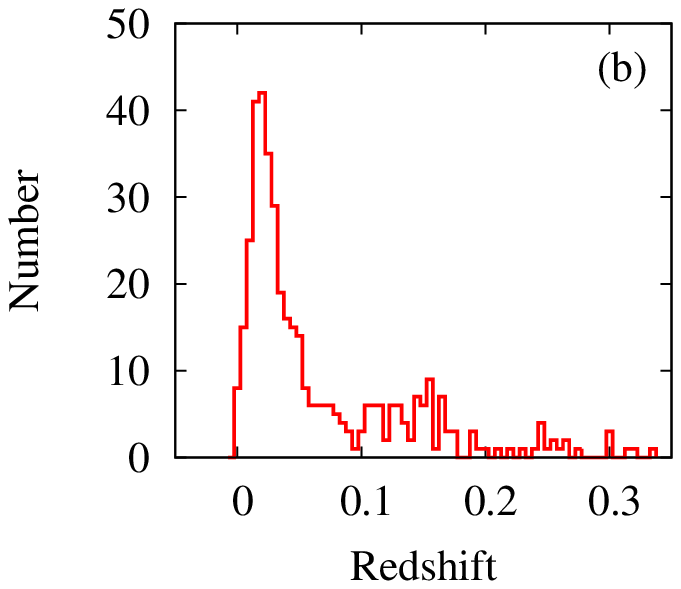}
  \end{minipage}
\caption{
  (a):Difference between redshifts from PAH emission, Br$\alpha$ emission, and literature ($z_{\rm spec}$).
  Our redshifts, $z$(PAH) and $z$(Br$\alpha$), are underestimated by $\Delta z\sim$0.004-0.005.
  (b): Redshift distribution of our sample.
}\label{fig:zs}
\end{figure*}

We investigate the difference between redshifts derived from PAH and Br$\alpha$, and those from literature. 
Fig.\ref{fig:zs}a shows that our redshifts are underestimated by $\Delta z\sim$ 0.004-0.005 and the variance of $\sigma(z)\sim$0.003.
This could be owing to the wavelength calibration or the registration of the world coordinate system (WCS) in the reduction processes. 
As the spectra were taken with a slit-less observation\footnote{Because the slit mask is as large as 1' $\times$ 1', it is equivalently a slit-less observation.}, uncertainty in the WCS directly affects the wavelength calibration.
This underestimation cannot be negligible, especially at $z < 0.01$. 
Hence, we did not use redshifts measured in our work. This leads five galaxies with no redshift, and hence no distance information. 

The figure also shows that the difference between $z$(PAH) and $z$(Br$\alpha$). The mean and standard deviation is $\overline{\Delta z}=-0.0007$ and $\sigma (\Delta z)=0.002$, respectively. 
Considering that one pixel in the spectral images corresponds to $\Delta z$=0.002-0.003, this difference is small enough.
In other words, the non-linear term of wavelength-calibration issue is negligible. 


We show the redshift distribution of our sample in Fig.\ref{fig:zs}b.
We can see that the redshift is distributed mostly at $z_{\rm spec}\sim$0.02 and up to at $z_{\rm spec}\sim$0.34. 

\subsection{Luminosity distribution}
\begin{figure}
 \begin{center}
   \FigureFile(90,){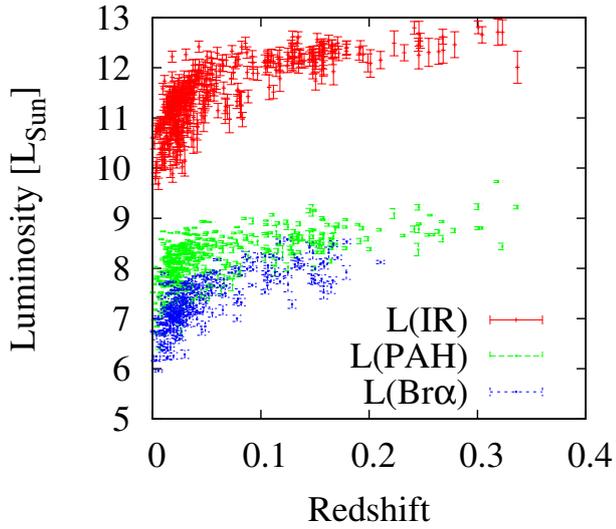} 
 \end{center}
\caption{Distribution of PAH 3.3 $\mu$m (green), Br$\alpha$ (blue), and infrared luminosity (red) against the redshift.}\label{fig:zslumi}
\end{figure}

In Fig.\ref{fig:zslumi} we show PAH 3.3 $\mu$m, Br$\alpha$ and infrared luminosity against the redshift.
Among our whole sample, 70\% have $L$(IR) $\rm > 10^{11}L_{\odot}$; 179 (43\%) and 111 (27\%) out of 412 galaxies are LIRGs and ULIRGs.
Specifically, 90\% of the galaxies at $z_{\rm spec} > 0.1$ (86 out of 95) are ULIRGs.
The PAH and Br$\alpha$ luminosity are roughly three and four orders of magnitudes lower than the $L$(IR), respectively.



\subsection{Galaxy type}
We classified our sample into five categories: SF, AGN, LINER, composite and unknown. 
We referred the SIMBAD and the Sloan digital sky survey (SDSS) data release 12. 
By using the SIMBAD, Seyfert (Sy), radio galaxies (rG), AGN, QSO and possible quasar were regarded as AGN, LINER was regarded as LINER, and HII galaxy and starburst galaxy were regarded as SF. 
From the SDSS catalogue, we used the ``BPT'' parameter, which classified sample into the above five categories.
Galaxies with a BPT parameter of ``Seyfert/LINER'' were classified as Seyfert.
In case the SIMBAD and the SDSS disagree, we prioritised the former.
In total our catalogue has 59 SF, 101 AGN, 50 LINER, 20 composite, 182 unknown (no information) galaxies. 


\clearpage
\begin{table}
  \rotatebox{90}{
    \begin{minipage}{\textheight}
      \centering
      \caption{PAH 3.3 $\mu$m, Br$\alpha$, and $L$(IR) catalogue of local galaxies observed with {\it AKARI}. This is a part of the catalogue. The full catalogue will be available from {\it AKARI} archive web page.
      \label{tab:cat}}
      \begin{tabular}{ccccccccccccccc}
        \hline
        Name & RA(J2000) & Dec(J2000) & $z_{\rm spec}$ & $D_{L}$ &{\it w1rsemi}\footnote{Semi-major axis of the elliptical aperture used to measure source in the {\it WISE} 3.4 $\mu$m band.} &Scale\footnote{$F(3.4)_{\it WISE}/F(3.4)_{\it AKARI}$ ratio use to scale line flux.} &log$L$(IR) &log$L$err(IR) &$\rm FISflag$\footnote{0:no blend, 1:sources are blended in the FIS all sky map} & type &Programme\\
        & & & & (Mpc) & (arcsec) & & (L$_{\odot}$) & & & & &\\
        \hline
2MFGC12572 & 234.04889 & -5.39759 & 0.0270 & 118.00 & 22.35 & 1.12 & 11.46 & 0.12 & 0 & LINER & AGNUL$\_$P3 \\
5MUSES171 & 241.16930 & 55.56920 & 0.0780 & 353.66 & 8.98 & 1.11 & 11.33 & 0.14 & 0 & Composite & AMUSE \\
AGC221050 & 192.55768 & 7.57904 & 0.0383 & 169.00 & 13.97 & 1.13 & 11.48 & 0.12 & 0 & SF & MSAGN \\
Arp148 & 165.97421 & 40.85007 & 0.0345 & 151.75 & 13.45 & 1.09 & 11.63 & 0.11 & 0 & Unknown & AGNUL$\_$P3 \\
CGCG049-057 & 228.30526 & 7.22500 & 0.0130 & 56.22 & 16.05 & 1.06 & 11.27 & 0.11 & 0 & Unknown & AGNUL$\_$P3 \\
ESO0255-IG007 & 96.84040 & -47.17670 & 0.0401 & 177.06 & 12.04 & 1.10 & 11.85 & 0.11 & 0 & Unknown & GOALS,DTIRC \\
IC0836 & 193.97508 & 63.61233 & 0.0092 & 39.66 & 38.48 & 1.25 & 10.16 & 0.12 & 0 & Unknown & MSAGN \\
IIIZw035 & 26.12708 & 17.10139 & 0.0274 & 119.80 & 11.75 & 0.94 & 11.59 & 0.11 & 0 & Unknown & GOALS \\
IRAS01173+1405 & 20.01130 & 14.36190 & 0.0312 & 136.93 & 14.21 & 1.04 & 11.64 & 0.11 & 0 & LINER & AGNUL$\_$P3 \\
        \hline
        \\
      \end{tabular}
      \begin{tabular}{cccccc | cccccc}
        \hline
        \multicolumn{6}{c}{PAH}  & \multicolumn{6}{c}{Br$\alpha$}\\
        $z$ & $zerr$ & Flux& Fluxerr & EW & EWerr & $z$ & $zerr$ & Flux & Fluxerr & EW & EWerr \\
        & & ($\rm erg/s/cm^{2}$) & ($\rm erg/s/cm^{2}$)& ($\mu$m) & ($\mu$m) & & & ($\rm erg/s/cm^{2}$)& ($\rm erg/s/cm^{2}$)& ($\mu$m) &($\mu$m)\\
        \hline
0.0248 & 0.0006 & 25.47 & 1.35 & 0.0553 & 0.0029 & 0.0234 & 0.0007 & 1.07 & 0.24 & 0.0035 & 0.0008 \\
0.0663 & 0.0011 & 3.20 & 0.31 & 0.0579 & 0.0057 & 0.0610 & 0.0004 & 0.26 & 0.04 & 0.0086 & 0.0012 \\
0.0347 & 0.0002 & 37.67 & 1.17 & 0.1188 & 0.0021 & 0.0350 & 0.0001 & 3.78 & 0.13 & 0.0185 & 0.0004 \\
0.0309 & 0.0003 & 27.86 & 0.84 & 0.0980 & 0.0030 & 0.0322 & 0.0002 & 2.35 & 0.12 & 0.0122 & 0.0006 \\
0.0096 & 0.0004 & 15.23 & 0.54 & 0.0439 & 0.0016 & 0.0105 & 0.0005 & 1.67 & 0.16 & 0.0080 & 0.0008 \\
0.0337 & 0.0002 & 66.34 & 1.28 & 0.1605 & 0.0031 & 0.0349 & 0.0001 & 7.05 & 0.21 & 0.0225 & 0.0007 \\
0.0109 & 0.0006 & 27.06 & 1.16 & 0.0387 & 0.0013 & 0.0077 & 0.0009 & 4.76 & 0.47 & 0.0116 & 0.0011 \\
0.0204 & 0.0004 & 9.52 & 0.49 & 0.0469 & 0.0019 & 0.0183 & 0.0010 & 1.79 & 0.21 & 0.0129 & 0.0014 \\
0.0227 & 0.0004 & 40.89 & 1.43 & 0.1182 & 0.0041 & 0.0243 & 0.0001 & 5.22 & 0.14 & 0.0142 & 0.0004 \\
\hline
      \end{tabular}
    \end{minipage}
  }
\end{table}
\clearpage

\section{Discussion}
\label{discuss}
In the previous section, we produced a catalogue of PAH 3.3 $\mu$m and Br$\alpha$ flux, together with the infrared luminosity. 
In this section we investigate relationships between these values. 

\subsection{L(PAH) vs $L$(IR)}
In Fig.\ref{fig:lirpah}a we show a ratio between the $L$(PAH) and infrared luminosity against the infrared luminosity. 
At $\rm log\, L$(IR)/L$_{\odot} < 11$, we see that the $L(\rm{PAH})$/$L$(IR) ratio is constant, $\rm \sim 10^{-3}$. 
At the higher $L$(IR), in contrast, this ratio decreases with the $L$(IR) and it is down to $\rm \sim 10^{-4}$ at $\rm log\,$ $L$(IR)/L$_{\odot} \sim 12.5$, consistent with e.g. \citet{Yamada13}.

Four optical types are colour-coded in the figure: SF including composite ones (blue), LINER (green), AGN (red), and unknown (grey).
We show median values with 68\% range in each 0.7 dex log LIR bin in Fig.\ref{fig:lirpah}c.
We found no strong dependence of $L(\rm{PAH})$/$L$(IR) on the galaxy type; any types of galaxies at higher $L$(IR) have lower $L(\rm{PAH})$/$L$(IR). 

In Fig.\ref{fig:lirpah}e we show the same figure for galaxies with $F_{\nu}(90)/F_{\nu}(140)>1$(blue) and $F_{\nu}(90)/F_{\nu}(140)<1$(red).
Only galaxies with $z_{\rm spec}<0.1$ are plotted.
The $F_{\nu}(90)/F_{\nu}(140)$ ratio reflects the dust temperature and the ratio of unity corresponds to $T_{dust}\sim$45 K.
We can see that galaxies with higher $F_{\nu}(90)/F_{\nu}(140)$ ratio tend to distribute at higher $L$(IR).
Nonetheless, at fixed $L$(IR), we found no dependence of $L(\rm{PAH})$/$L$(IR) on the $F_{\nu}(90)/F_{\nu}(140)$ ratio.



\subsection{L(Br$\alpha$) vs $L$(IR)}
\begin{figure*}[ht]
  \centering
   \begin{minipage}{0.49 \hsize}
     \FigureFile(80,){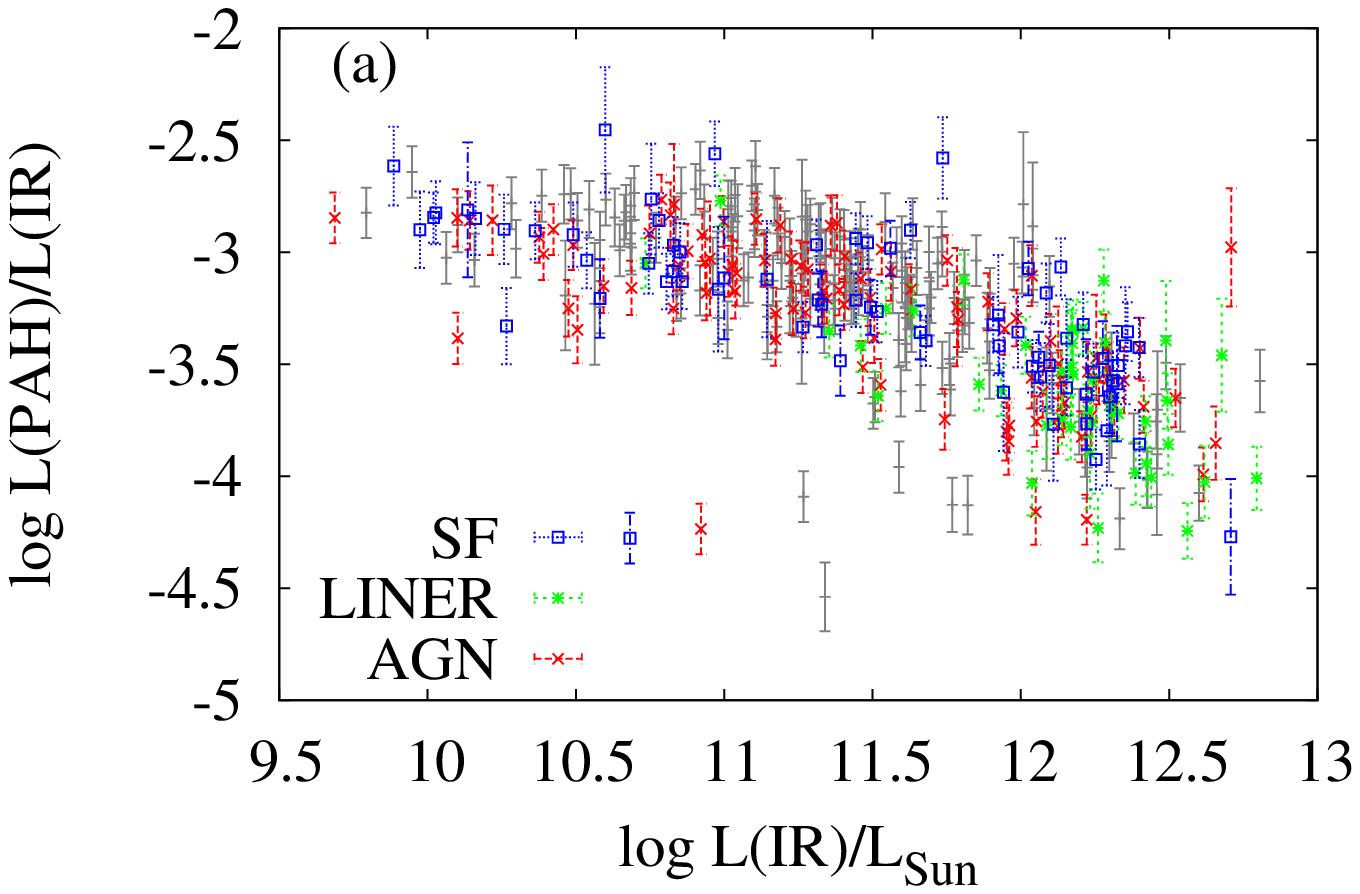}
   \end{minipage}
   \begin{minipage}{0.49 \hsize}
     \FigureFile(80,){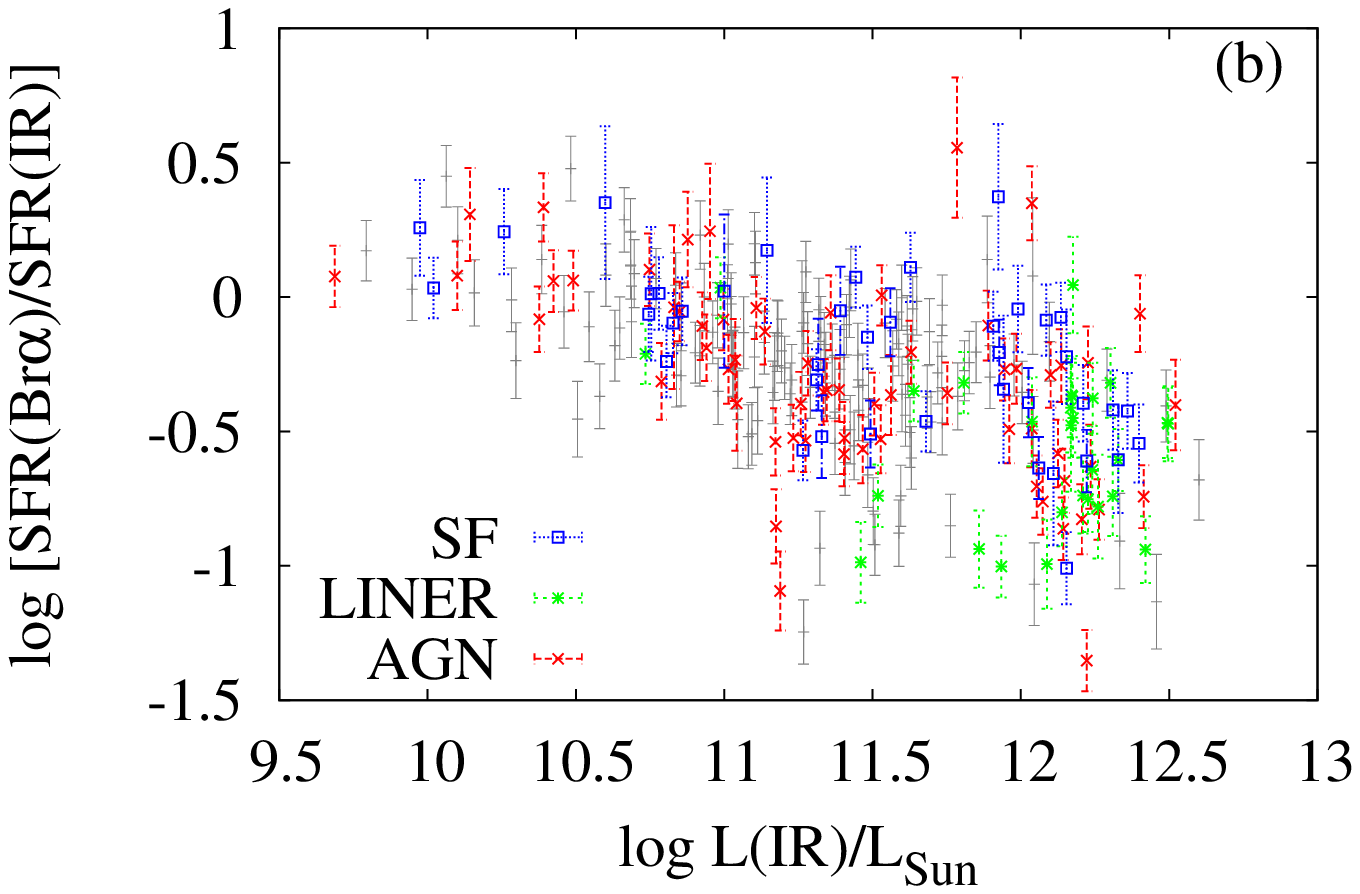}
   \end{minipage}

   \begin{minipage}{0.49 \hsize}
     \FigureFile(80,){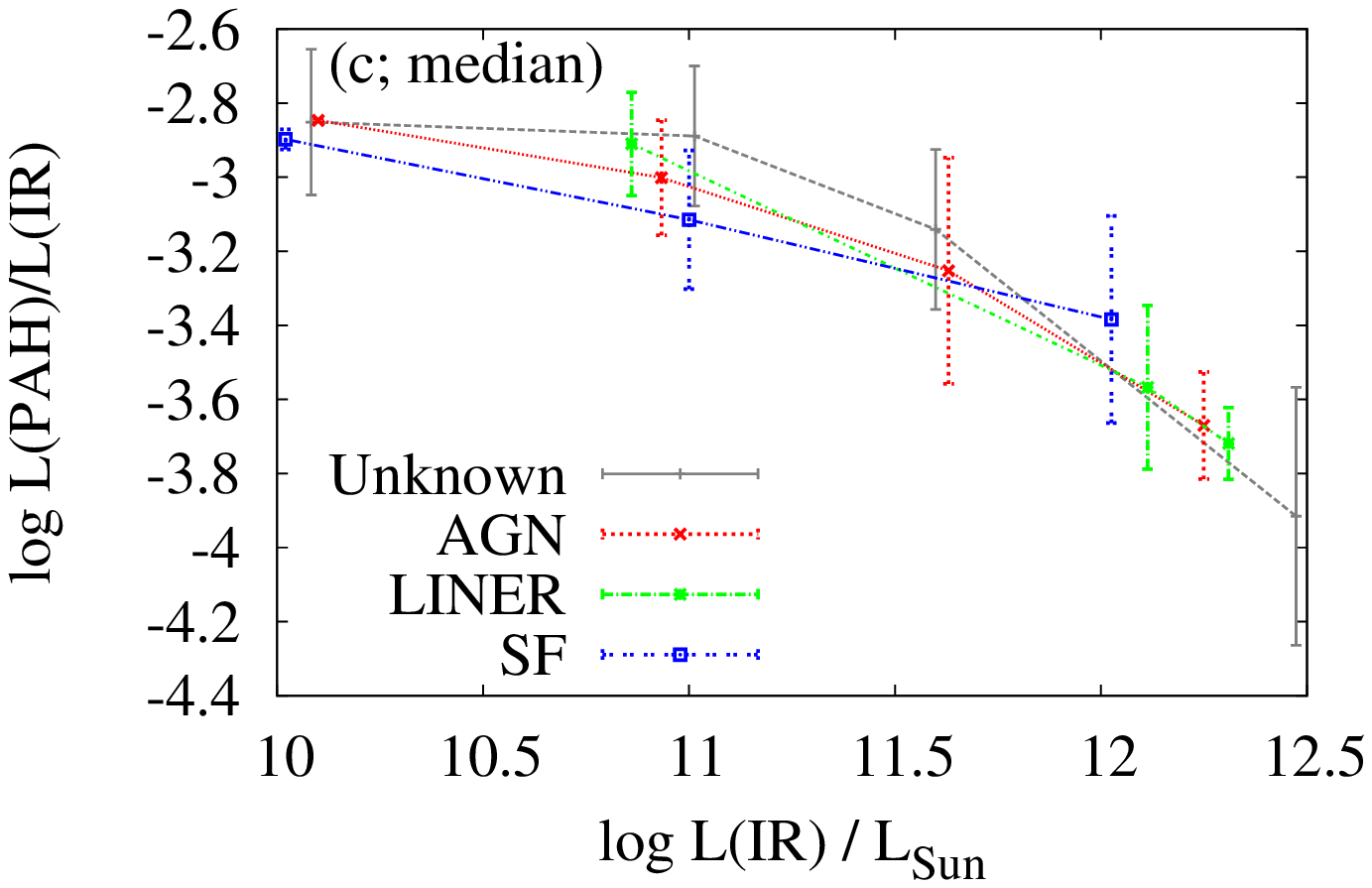}
   \end{minipage}
   \begin{minipage}{0.49 \hsize}
     \FigureFile(80,){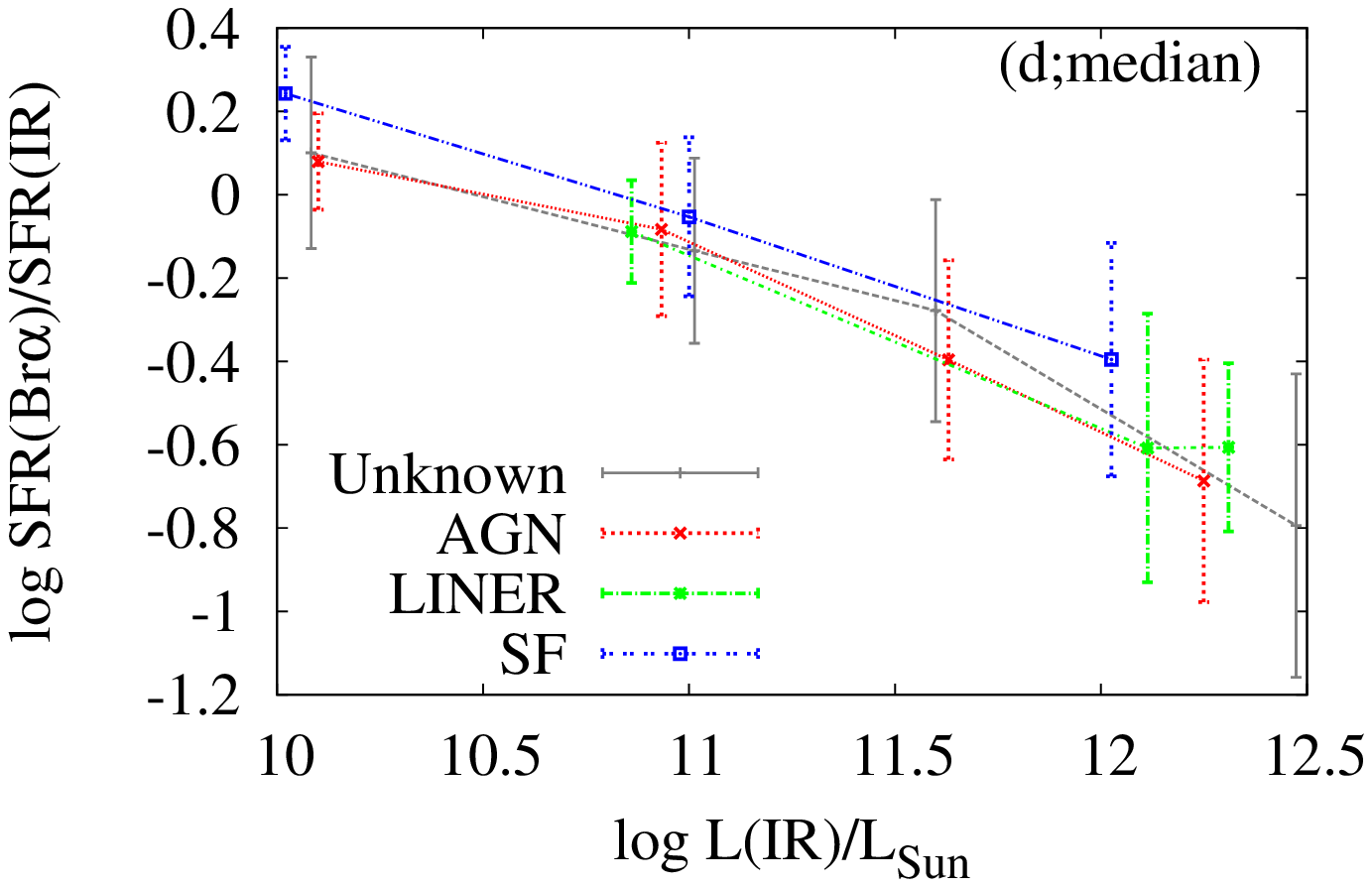}
   \end{minipage}
   
   \begin{minipage}{0.49 \hsize}
     \FigureFile(80,){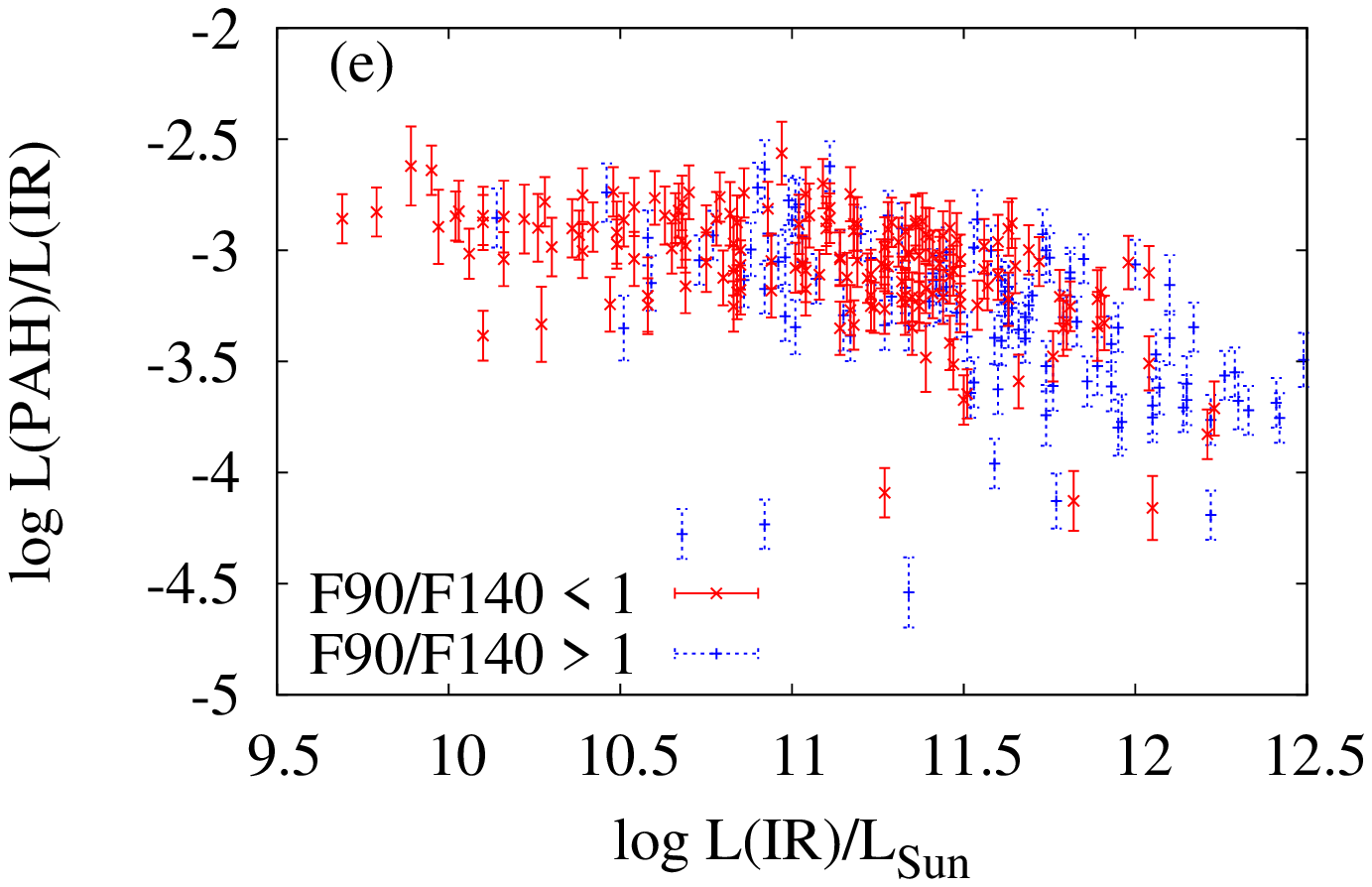}
   \end{minipage}
   \begin{minipage}{0.49 \hsize}
     \FigureFile(80,){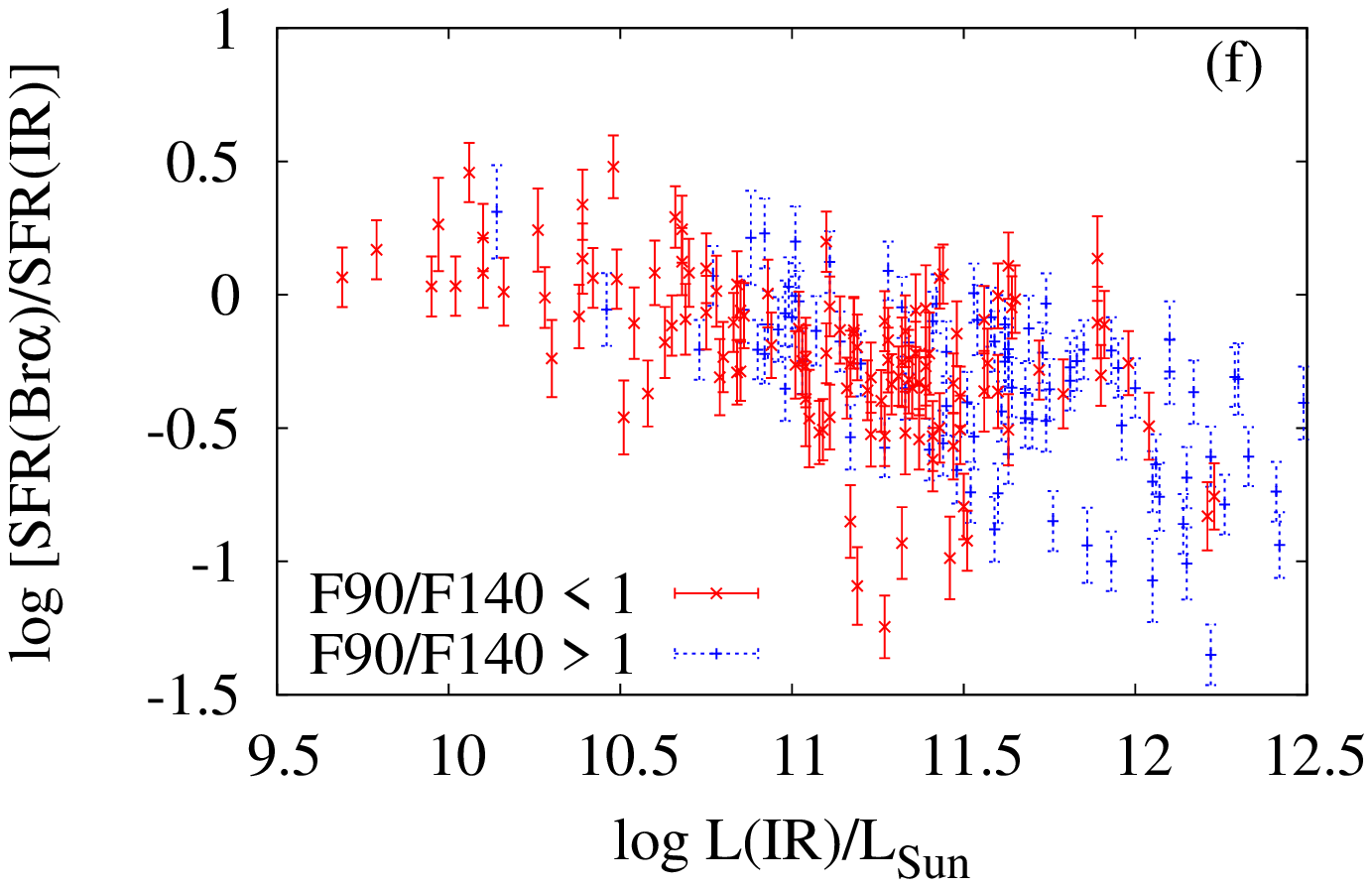}
   \end{minipage}
   
\caption{(a): Ratio between PAH 3.3 $\mu$m luminosity to $L$(IR) against the $L$(IR).
  The sample is divided into four type: SF (blue), LINER (green), AGN (red), and unknown (grey).
  (b): Ratio between SFR(Br$\alpha$) to SFR(IR) against the $L$(IR).
  (c): Medians of the $L(\rm{PAH})$/$L$(IR) ratio for each galaxy type.
  (d): Same as the panel (c) but for SFR$(\rm{Br\alpha})$/SFR(IR) ratio. 
  (e): Same as the panel (a) but the sample is divided with $F_{\nu}(90)/F_{\nu}(140)$. 
  (f): Same as the panel (b) but the sample is divided with $F_{\nu}(90)/F_{\nu}(140)$. 
\label{fig:lirpah}}
\end{figure*}

In Fig.\ref{fig:lirpah}b we compare SFR derived from the Br$\alpha$ emission and that from the $L$(IR), as a function of $L$(IR). 
For estimating SFR we used the following equations \citep{Kennicutt1994,Kennicutt1998apj}.
\begin{eqnarray}
  \rm{SFR}&=&7.94\times10^{-42}\,L\rm{\left( H\alpha \right)[erg/s]}\nonumber \\
  &=&2.85\times10^{-40}\,L\rm{\left( Br\alpha \right)[erg/s]} \nonumber \\
  &=&1.10\times10^{-6}\,L\rm{\left( Br\alpha \right)[L_{\odot}]},
\end{eqnarray}
%
%
\begin{eqnarray}
  \rm{SFR}=1.72\times 10^{-10} \, \rm {L(IR) \,[L_{\odot}]},
\end{eqnarray}
where we assume H$\alpha$/Br$\alpha$ ratio of 36, according to the case B with T=10000 K and $n_{e}$=10$^{2-4}$[cm$^{-3}$] \citep{Hummer87}, and the Salpeter IMF. 
In Fig.\ref{fig:lirpah}b we can see that the SFR(Br$\alpha$) is consistent with SFR(IR) at the log $L$(IR)/L$_{\odot}$$\sim$10.5. 
Meanwhile, at the higher $L$(IR), the SFR(Br$\alpha$)/SFR(IR) ratio gradually decreases with the $L$(IR) and it is down to 0.1 at log $L$(IR)/$L_{\odot}$ $\sim$12. 

We also show medians for each galaxy type in Fig.\ref{fig:lirpah}d. 
The SFR(Br$\alpha$)/SFR(IR) ratio also has no strong dependence on the galaxy optical type.
The relation is similar to that of the $L(\rm{PAH})$/$L$(IR) ratio.
Both ratios drop $\sim$1 dex from $\rm log L(IR)=10$ to $\rm log L(IR)=12.5$, and their variance is $\sigma_{68\%}\sim$0.2 dex. 

In Fig.\ref{fig:lirpah}f we show the same figure for galaxies at $z_{\rm spec}<0.1$ separated with the $F_{\nu}(90)/F_{\nu}(140)$ ratio.
Same as the $L(\rm{PAH})$/$L$(IR) ratio, we found no strong dependence of SFR(Br$\alpha$)/SFR(IR) on the $F_{\nu}(90)/F_{\nu}(140)$ ratio.

\subsection{PAH vs Br$\alpha$}
\begin{figure*}
  \centering
    \FigureFile(90,){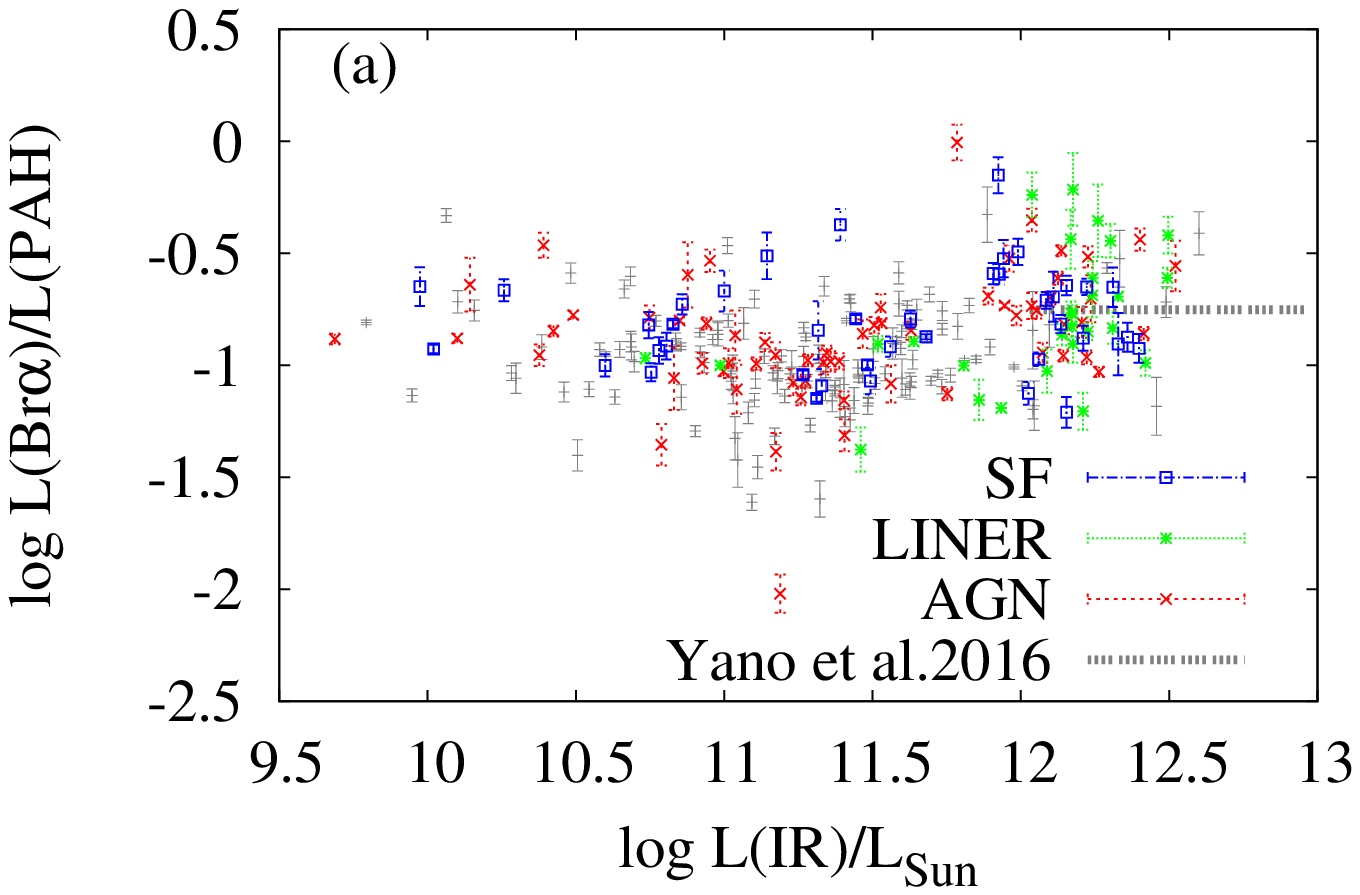}
    \FigureFile(90,){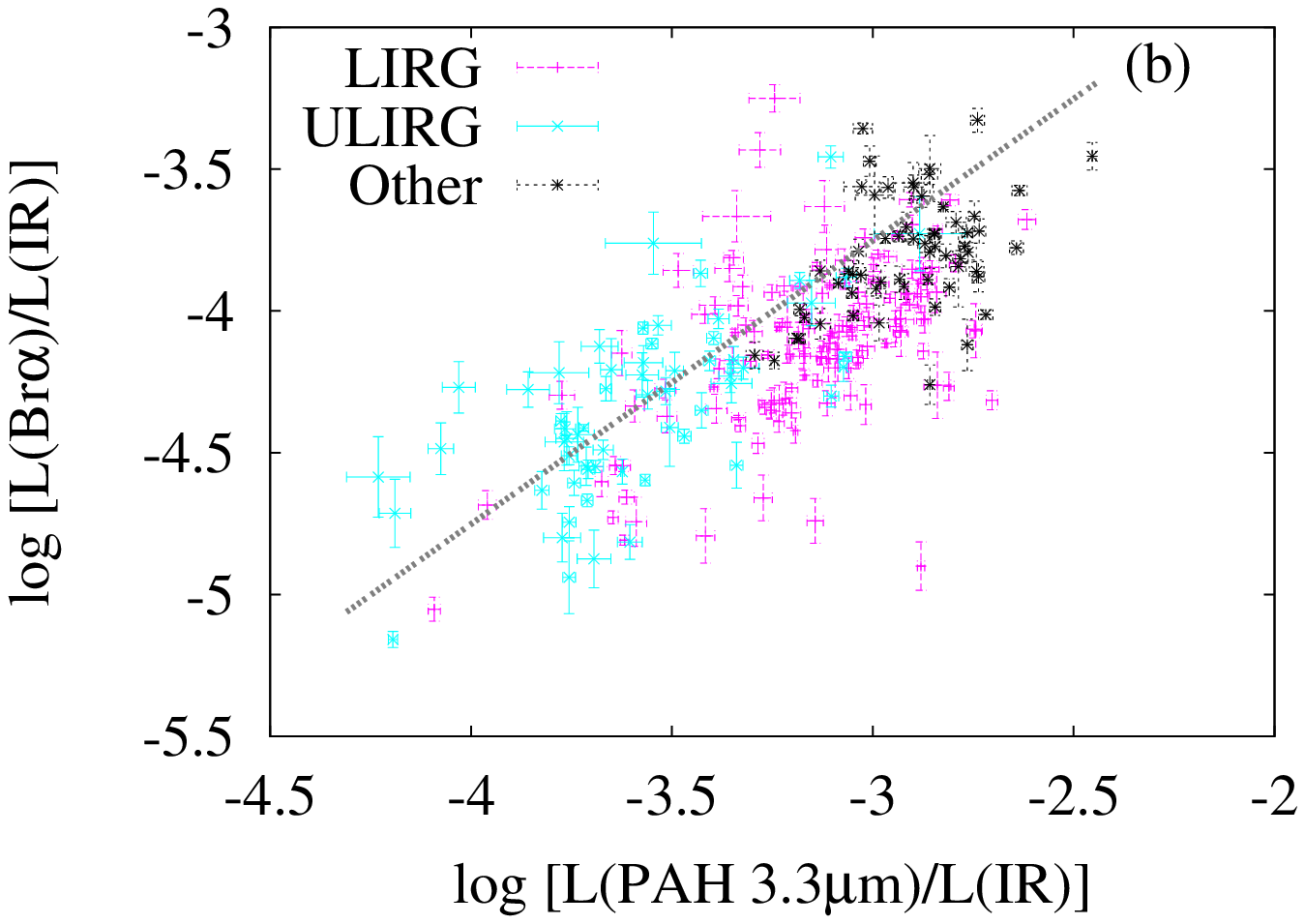}
\caption{Comparison between Br$\alpha$ and PAH luminosity.
  (a): Ratio between them against the $L$(IR). A ratio estimated for local ULIRGs in \citet{Yano16} is also shown in the grey broken line. The colour code is the same as Fig.5a.
  (b): Both luminosity are normalised with their $L$(IR).
  They are classified with their $L$(IR): LIRGs (magenta), ULIRGs (cyan), and fainter than LIRGs (black).
  The grey broken line indicates, again, the ratio estimated in \citet{Yano16}.
}\label{fig:lirpahbra}
\end{figure*}

To understand the behaviour of PAH and Br$\alpha$ emission lines in detail, we show $L(\rm{Br\alpha})$ to $L(\rm{PAH})$ ratio against the $L$(IR) in Fig.\ref{fig:lirpahbra}a.
The $L(\rm{Br\alpha})$/$L(\rm{PAH})$ ratio is consistent with that reported in Yano et al.(2016; grey broken line) for ULIRGs whereas at around $L$(IR)$\sim 10^{11}L_{\odot}$ the ratio is $>$3 times less than that of ULIRGs.
The decrease of L(PAH) as the $L$(IR) increase is faster than the decrease of Br$\alpha$ along the $L$(IR).
%
We see only a few galaxies in the bottom side of the figure, which could be a selection bias; as Br$\alpha$ emission is significantly fainter than the PAH 3.3 $\mu$m emission, we may not detect the Br$\alpha$ emission from galaxies in this region. 
Galaxies with no detection of Br$\alpha$ emission have log L(IR) of 10$\rm ^{10-12.5}L_{\odot}$.
Assuming their Br$\alpha$ flux is $<$ 0.5$\times$10$^{-14}$ erg/s/cm$^{2}$ (see Fig.\ref{fig:ewflxhist}a), half of them would distribute at log $L$(IR)/L$_{\odot}$$<$12 and log $L\rm(Br\alpha)$/$L$(PAH)$<$ -1.3, and the other half would distribute at log $L$(IR)/L$_{\odot}>$12 and log $L\rm(Br\alpha)$/$L$(PAH)$<$ 0. 

In Fig.\ref{fig:lirpahbra}b, we compare $L(\rm{Br\alpha})$/$L$(IR) and $L(\rm{PAH})$/$L$(IR) of LIRGs, ULIRGs, and the other sample.  
We can see again from this figure that galaxies with higher $L$(IR) have lower $L(\rm{Br\alpha})$/$L$(IR) and $L(\rm{PAH})$/$L$(IR) ratios.
The figure shows a clear correlation, where the correlation coefficient is 0.703.  
The clear correlation between these ratios implies that the origins of low $L\rm (PAH)$/$L$(IR) and $L\rm(Br\alpha)$/$L$(IR) are the same, or at least, related to each other. 
The figure also shows that our LIRG sample has lower $L(\rm{Br\alpha})$/$L$(IR) ratio than that reported from Yano et al.(2016) for ULIRG sample. It implies destruction of PAH molecules in ULIRGs (see also section 5.4).

\subsection{Interpretation}
In the previous sections, we found that both $L$(PAH)/$L$(IR) and $L$(Br$\alpha$)/$L$(IR) are lower at higher $L$(IR), and they correlate with each other.
The relative weakness of the two ratios and their correlation can be interpreted in the following four ways. 

(a) Large extinction of PAH and/or Br$\alpha$ emissions: Despite the long wavelength of these emissions, dust extinction could not be negligible.
In the galactic extinction curve the $k_{\lambda}$ is 0.286 at 3.29 $\mu$m and 0.253 at 4.052 $\mu$m, respectively.
It is 0.08$\sim$0.09 times as much as the Rv.
Hence, if the galaxies have strong dust extinction of Av$\sim$30 mag, both $L$(PAH)/$L$(IR) and $L$(Br$\alpha$)/$L$(IR) could be lower by one order of magnitude.

If this is the case we have to note two things; First, a significant amount of dust should be in or around the PDR because PAH molecules are in the PDR.
Second, the most energy is likely to come from the very centre of galaxies because extremely high dust attenuation in the entire region is not likely.

(b) Destruction of PAH molecules: Harsh radiation from an active galactic nucleus (AGN) or strong starburst may destroy them \citep{Nordon12,Murata14}.
This can explain only lower $L$(PAH)/$L$(IR) ratios. Since our ULIRG sample shows lower $L$(PAH)/$L$(IR) ratios than those of LIRG sample, destruction of PAH molecules may be more effective in ULIRGs. 
Nonetheless, the correlation between $L$(Br$\alpha$)/$L$(IR) and $L$(PAH)/$L$(IR) ratios indicates other phenomenon to make $L$(Br$\alpha$)/$L$(IR) and $L$(PAH)/$L$(IR) ratios lower. 

(c) Lack of UV photon ionising hydrogen gas and exciting PAH molecules: If UV photons produced from young stars are absorbed by dust before it ionises gas and excites PAH molecules, both the hydrogen recombination lines and PAH emission is weakened \citep{Valdes05,Murata14,Yano16}, which results in the drop of $L$(PAH)/$L$(IR) and the $L(Br\alpha)$/$L$(IR).

To explain the observed (minimum) SFR(Br$\alpha$)/SFR(IR)$<$0.1 based on this, more than 90\% of ionising photons should be absorbed.
Similar to our study, \citet{Valdes05} studied the fraction of ionising photons absorbed by dust for local LIRGs using Br$\gamma$ and Pa$\alpha$ lines.
They showed that the fraction should be $\sim$80\% for explaining observed ratios between $L(\rm Br\gamma)$ or $L(\rm Pa\alpha)$ to $L$(IR).
\citet{Yano16} also suggested that the fraction is, on average, $\sim$45\% for 13 local HII ULIRGs using a ratio between Br$\alpha$ and $L$(IR).
This value is lower than ours, which could be due to their small sample size.

(d) AGN contribution to $L$(IR):
If AGNs significantly contribute their energy to the $L$(IR) and do not to the $L\rm (Br \alpha)$ and $L$(PAH), we expect a lower $L\rm (Br\alpha)$/$L$(IR) and $L$(PAH)/$L$(IR) ratios than those of normal star-forming galaxies.

We found no strong dependence of $L(\rm{PAH})$/$L$(IR) and $L\rm (Br\alpha)$/$L$(IR) on galaxy type. It implies the AGN contribution is not a dominant cause. 
This is inconsistent with \citet{Yano16} who showed that the $L(\rm{Br}\alpha$)/$L$(IR) ratio depends on the galaxy type; the ratio of HII galaxies is significantly higher than that of  LINER and Seyfert galaxies. 
This inconsistency could be due to the sample size; we have as many as 230 classified galaxies while \citet{Yano16} have only $\sim$30 galaxies.
Hence we conclude that our result is more robust. 


Nonetheless, we have to consider obscured AGNs. 
\citet{Imanishi10} showed that even HII galaxies could have an AGN.
If it significantly contributes the energy into $L$(IR), both $L(\rm{PAH})$/$L$(IR) and $L(\rm{Br}\alpha$)/$L$(IR) could be lower.
\citet{Symeonidis16} showed that the intrinsic far-infrared emission from QSOs is higher than previously known. 
They implied that contribution from a powerful AGN to any broad band flux cannot be neglected for estimating the SFR of the host galaxy. 
Hence, although we found no dependence of $L(\rm{PAH})$/$L$(IR) and $L(\rm{Br}\alpha$)/$L$(IR) on galaxy type, we cannot reject the scenario (d). 

\section{Summary}
In this work we presented a catalogue of PAH 3.3 $\mu$m, Br$\alpha$, and infrared luminosity of local galaxies observed with {\it AKARI}. 
We measured PAH 3.3 $\mu$m and Br$\alpha$ flux from the {\it AKARI}/IRC 2-5 $\mu$m spectra. 
We detected PAH emission from 412 galaxies and Br$\alpha$ emission from 264 galaxies.
Among them, we measured infrared luminosity of 380 galaxies using the {\it AKARI}-far-infrared-all-sky-survey data. 

Most of our sample have PAH flux of $\rm 1-100 \times 10^{-14}\, erg\, s^{-1}\, cm^{-2} $, and Br$\alpha$ flux of $\rm 1-10 \times 10^{-14} erg\, s^{-1}\, cm^{-2} $, and infrared luminosity of $\rm 10^{10 - 12}\, L_\odot$. 
The redshift range of our sample is $z_{spec} \sim 0.002-0.3$. 

Using this catalogue, we investigated relations between the three kinds of luminosity.
We found that at log $L$(IR) $>$ $10^{11}\rm L_\odot$ both PAH 3.3 $\mu$m and Br$\alpha$ luminosity show a relative weakness compared with the $L$(IR).
We found that their relative weakness is not strongly dependent on galaxy type and dust temperature. 
We attributed the relative weakness to
a.) extremely strong dust attenuation that reduces even near-infrared light,
b.) destruction of PAH molecules due to harsh radiation,
c.) a lack of UV photons that excite PAH molecules or ionise hydrogen gas,
or d.) none-SF components such as AGNs contributing to the $L$(IR). 
Although we cannot determine how much they contribute to the relative weakness found in our work, a clear correlation between $L(\rm{PAH})$/$L$(IR) and $L(\rm{Br\alpha})$/$L$(IR) implies that the origins of their relative weakness are related with each other. 


\bigskip


We thank Mr. Arihiro Kamada, Mr. Kengo Yusou, and Ms. Kanae Ohi.
They worked with us on producing the catalogue as a JAXA internship student. 

This research is based on observations with AKARI, a JAXA project with the participation of ESA. 

This research has made use of the SIMBAD database,
operated at CDS, Strasbourg, France \citep{Wagner00}.

 This research has made use of the NASA/IPAC Extragalactic Database (NED) which is operated by the Jet Propulsion Laboratory, California Institute of Technology, under contract with the National Aeronautics and Space Administration.

This work was financially supported in part by the Grants-in-Aid for Scientific Research (KAKENHI, 26247030).

\end{document}